\documentstyle[12pt,epsf]{article}
\def\beq{\begin{equation}}
\def\eeq{\end{equation}}
\def\ap#1#2#3 {Ann. Phys. (NY) {\bf#1} (19#2) #3}
\def\apj#1#2#3 {Astrophys. J. {\bf#1} (19#2) #3}
\def\apjl#1#2#3 {Astrophys. J. Lett. {\bf#1} (19#2) #3}
\def\app#1#2#3 {Acta. Phys. Pol. {\bf#1} (19#2) #3}
\def\ar#1#2#3 {Ann. Rev. Nucl. Part. Sci. {\bf#1} (19#2) #3}
\def\cpc#1#2#3 {Computer Phys. Comm. {\bf#1} (19#2) #3}
\def\err#1#2#3 {{\it Erratum} {\bf#1} (19#2) #3}
\def\ib#1#2#3 {{\it ibid.} {\bf#1} (19#2) #3}
\def\jmp#1#2#3 {J. Math. Phys. {\bf#1} (19#2) #3}
\def\ijmp#1#2#3 {Int. J. Mod. Phys. {\bf#1} (19#2) #3}
\def\jetp#1#2#3 {JETP Lett. {\bf#1} (19#2) #3}
\def\jpg#1#2#3 {J. Phys. G. {\bf#1} (19#2) #3}
\def\mpl#1#2#3 {Mod. Phys. Lett. {\bf#1} (19#2) #3}
\def\nat#1#2#3 {Nature (London) {\bf#1} (19#2) #3}
\def\nc#1#2#3 {Nuovo Cim. {\bf#1} (19#2) #3}
\def\nim#1#2#3 {Nucl. Instr. Meth. {\bf#1} (19#2) #3}
\def\np#1#2#3 {Nucl. Phys. {\bf#1} (19#2) #3}
\def\pcps#1#2#3 {Proc. Cam. Phil. Soc. {\bf#1} (#2) #3}
\def\pl#1#2#3 {Phys. Lett. {\bf#1} (19#2) #3}
\def\prep#1#2#3 {Phys. Rep. {\bf#1} (19#2) #3}
\def\prev#1#2#3 {Phys. Rev. {\bf#1} (19#2) #3}
\def\prl#1#2#3 {Phys. Rev. Lett. {\bf#1} (19#2) #3}
\def\prs#1#2#3 {Proc. Roy. Soc. {\bf#1} (19#2) #3}
\def\ptp#1#2#3 {Prog. Th. Phys. {\bf#1} (19#2) #3}
\def\ps#1#2#3 {Physica Scripta {\bf#1} (19#2) #3}
\def\rmp#1#2#3 {Rev. Mod. Phys. {\bf#1} (19#2) #3}
\def\rpp#1#2#3 {Rep. Prog. Phys. {\bf#1} (19#2) #3}
\def\sjnp#1#2#3 {Sov. J. Nucl. Phys. {\bf#1} (19#2) #3}
\def\spj#1#2#3 {Sov. Phys. JEPT {\bf#1} (19#2) #3}
\def\spu#1#2#3 {Sov. Phys. Usp. {\bf#1} (19#2) #3}
\def\zp#1#2#3 {Zeit. Phys. {\bf#1} (19#2) #3}

\begin{document}
\begin{titlepage}
\begin{center}
{\Large \bf Theoretical Physics Institute \\
University of Minnesota \\}  \end{center}
\vspace{0.2in}
\begin{flushright}
TPI-MINN-97/25-T \\
UMN-TH-1606-97 \\
hep-th/9709137\\
\end{flushright}
\vspace{0.3in}
\begin{center}
{\Large \bf  Degenerate Domain Wall Solutions in  Supersymmetric
Theories\\}
\vspace{0.2in}
{\bf M.A. Shifman \\}
Theoretical Physics Institute, University of Minnesota, Minneapolis,
MN
55455 \\

\vspace{0.2cm}

and

\vspace{0.2cm}

{\bf M.B. Voloshin  \\ }
Theoretical Physics Institute, University of Minnesota, Minneapolis,
MN
55455 \\ and \\
Institute of Theoretical and Experimental Physics, Moscow, 117259
\\[0.2in]
\end{center}

\begin{abstract}

A family of degenerate domain wall configurations, partially preserving
supersymmetry, is discussed in a generalized Wess-Zumino model  with
two scalar superfields. We establish some general features inherent to
the models with continuously degenerate domain walls. For instance,
for purely real trajectories additional ``integrals of motion" exist.
The solution for the profile of the scalar fields for any wall belonging
to the family  is found in quadratures  for arbitrary ratio of the
coupling constants. For a special value of this ratio the solution
family is obtained explicitly, in terms of elementary functions. We
also discuss the threshold amplitudes for multiparticle production
generated by these solutions. New unexpected nullifications of the
threshold amplitudes are found.

\end{abstract}
\end{titlepage}

\section{Introduction}

Existence of several degenerate supersymmetric vacua is a generic
phenomenon in supersymmetric theories with scalar superfields. Moreover,
in many instances the vacuum manifold consists of several isolated
points. Thus, the possibility arises of domain wall configurations
interpolating between these vacua.

It has been recently shown$^{\cite{ds1,ds2}}$ that some of the domain
wall configurations in  (3+1) dimensional theories posses distinct
supersymmetric properties:\\

{\it i)} they generate a central extension of the $N=1$ superalgebra.
The wall tension is proportional to the central charge.  Due to the
non-renormalization theorem for the central charge this implies that
the wall energy density is exactly calculable, it is not renormalized
by the loop corrections.

{\it ii)} they preserve two out of four original supercharges (``N=1/2
supersymmetry") corresponding to the  minimal supersymmetry in
the (2+1) dimensional space tangential to the wall;

{\it iii)} the profile of the fields across these walls satisfies first
order differential equations analogous to the
Bogomolny-Prasad-Sommerfeld (BPS) equations$^{\cite{bps}}$ (the so
called ``BPS-saturated walls", or ``BPS walls"). These equations were
called the creek equations$^{\cite{cs}}$ because of a mechanical
interpretation \footnote{Previously the creek equations were considered
in different contexts in Refs. \cite{1} -- \cite{2}.}.\\

A non-vanishing central charge of the $N=1$ superalgebra  exists for any field
configuration interpolating between two distinct
vacua$^{\cite{cs}}$. Not every such configuration is BPS saturated,
however. Those domain walls that are BPS saturated posses peculiar
features following from the point {\it ii} above.  The BPS domain walls
have interesting properties even in the simplest theories, e.g. in the
Wess-Zumino model with one chiral superfield. Even more remarkable they
become in the theories with two or more chiral superfields interacting
with each other.

In the theories with $K $ scalar superfields generically there are $2^K$
degenerate vacuum states and, correspondingly, there are at least
$2^{K-1} \, (2^K-1)$ domain wall types$^{\cite{mv}}$. In Ref. \cite{ms}
it was shown that, quite typically,  some of these domain walls turn out
to be continuously degenerate. Collective coordinates exist
corresponding to a continuous deformation of the internal structure of
the wall. Varying these coordinates we change the wall structure leaving
the wall energy density intact.

In this work we investigate this phenomenon, both in general aspect and
in some simple examples. It will be argued that the continuous
degeneracy is related to the existence of additional ``integrals of
motion". We will explicitly find such an integral in a particular
two-field model. Using this result it turns out possible to obtain a
generic family of the domain wall solutions in this two-field model. For
arbitrary values of the coupling constants the solution is given in
terms of quadratures. For some specific values a closed-form solution in
terms of elementary functions exists. We take advantage of this explicit
solution to extract consequences for the high-order multiparticle
amplitudes at threshold.

The supersymmetric vacua are  determined by the extrema of the
superpotential $W(\phi_k)$,
\beq
\partial W /\partial \phi_k=0\, ,
\,\,\, k=1 \ldots K\, .
\label{svac}
\eeq
The general form of the
BPS-saturation equations for a static wall in which the fields $\phi_i$
depend only on the coordinate $z$ is
\beq
\partial_z  \phi_k = {\partial
W^\dagger \over \partial \phi_k^\dagger} \, e^{i \alpha}~~,
\label{bps}
\eeq
where $\alpha$ is a constant ($z$ independent) complex phase. Let
us assume that two solutions of the equations (\ref{svac}) are found,
$\{\phi\}_1$ and $\{\phi\}_2$, where the braces denote a set of all
scalar fields $\phi_k$. Denote the corresponding values of the
superpotential by
\beq
W_1 \equiv  W(\{\phi\}_1)\, , \,\,\,  W_2 \equiv
W(\{\phi\}_2)~~ .
\eeq
Without loss of generality one can assume that
$W_1 =0$ and $W_2$ is real and positive (this can be always achieved by
appropriate transformations of the superpotential). Then  the phase
$\alpha$ in Eq. (\ref{bps}) can be set equal to zero. If, additionally,
the superpotential is real for real values of the superfields, to
which we limit our investigation, then Eqs. (\ref{bps}) take the form
\beq
\partial_z  \phi_k = {\partial  W \over \partial \phi_k}\, .
\label{bpsr}
\eeq
Now, if one  interprets $z$ as ``time", the latter
equations have  a simple mechanical inter\-pretation$^{\cite{cs}}$: they
describe the flow of a very viscous fluid, whose inertia can be
neglected, in the potential relief $-W(\phi_k)$ from one extremum $-W_1$
of the relief, along a gradient line, to a lower extremum $-W_2$.
(Obviously, reflection of the $z$ direction is possible, in which case
the flow is described by the ``potential" $W$ rather than $-W$. Instead
of the reversal of the $z$ axis one can view this as a shift by $\pi$ of
the phase $\alpha$ in Eq. (\ref{bps}). For definiteness we use the
mechanical analogy within the conventions of Eq. (\ref{bpsr}).)  From
the fluid flow analogy it is clear that the necessary condition for the
existence of a solution of the equations (\ref{bpsr}) is that the
extrema $W_1$ and $W_2$ be of different height. One can take advantage
of a rich intuition one has in connection with the mechanical motion of
this type. To make this analogy more graphic we will sometimes denote
the derivative over $z$ as
$$ \partial_z\phi \rightarrow \dot\phi $$
in the cases where there is no menace of confusion. Correspondingly, the
quantities conserved along the given trajectory will be referred to as
integrals of motion. One integral of motion, energy, is well-known of
course; it is universal and has nothing to do with the specific
trajectories under consideration. We will be interested in additional
integrals of motion, specific to the creek equations.

The surface energy density of the BPS wall is given by the difference of
the superpotential at the two extrema,
\beq
\varepsilon = 2(W_2-W_1)~~.
\label{ener}
\eeq

Although at least some of the  results to be reported below seem to be
generic, we will make no attempt at formulating them in full generality.
Instead,  we will  focus on  a specific instructive  example:   a
two-field model$^{\cite{mv,ms}}$ with the scalar superfields $\Phi$ and
$X$ and  the superpotential
\beq
W(\Phi, \, X)= {m^2 \over \lambda} \, \Phi - {1 \over 3} \, \lambda
\,
\Phi^3 - \alpha \, \Phi \, X^2 \, .
\label{spot}
\eeq
Here $m$ is a mass parameter and $\lambda$ and $\alpha$ are
dimensionless coupling constants. It is  assumed that the phases of the
fields and $W$ are adjusted  in such a way that all parameters in the
superpotential are real and positive. Only occasionally we will digress
to more general models. Below the lowest component of $\Phi$ and $X$
will be denoted by $\phi$ and $\chi$, respectively.

Clearly, the superpotential (\ref{spot})  is not the most general form
of the superpotential in the  renormalizable (3+1) dimensional  models,
even given the freedom of redefining the fields. It contains, however,
sufficient features for a discussion of the non-trivial properties of
the domain walls we are interested in. \footnote{An equivalent model was
also considered in Ref. \cite{morris}, and a  similar model  of higher
order in the fields in Ref. \cite{etbb}, in connection with different
problems.}  The model with the superpotential (\ref{spot}) will be
referred to below as a minimal two-field model. In  terms of the scalar
components $\phi$ and $\chi$ of the respective superfields $\Phi$ and
$X$ the ``potential relief" $-W$ has its maximum $-W_1$ at $\phi=-
m/\lambda$,  $\chi=0$ and the minimum $-W_2$ at $\phi=m/\lambda$,
$\chi=0$. It also has two saddle points of equal height at $\phi=0$,
$\chi=\pm m/\sqrt{\lambda  \, \alpha}$. The BPS-saturated walls  exist
connecting the  maximum and the minimum, and also connecting either of
the saddle points with the maximum or the minimum. Moreover, all these
BPS-saturated configurations belong to one and the same family of
solutions, corresponding to the flow from the maximum to the minimum
with different starting conditions$^{\cite{ms}}$. All trajectories from
the family are real.

This is an ideal setting for establishing the existence of additional
integrals of motion. The one relevant to the model (\ref{spot})
is obtained in an explicit form. Certainly, given the additional
constraint, one can readily reconstruct the full family of solutions, in
quadratures. Further simplifications arise (a) for arbitrary values of
$\lambda$, $\alpha$ and the vanishing value of the integral of
motion;
(b) for $\lambda /\alpha = 4$ and arbitrary value of the integral of
motion. As a matter of fact, the first case was treated in Ref.
\cite{ms} where it was found
\beq
\phi(z)={m \over \lambda} \, \tanh \left ( {2 \, \alpha \over
\lambda}
\, m \, z \right
)~~, ~~~~\chi(z)=\pm {m \over \sqrt{\lambda \, \alpha} } \, \sqrt{ 1-
{2
\, \alpha
\over \lambda} } \, \, {\rm sech} \left ({2 \, \alpha \over \lambda} \,
m \, z \right
)~~.
\label{msol}
\eeq
Up till now, this was the only  non-trivial solution  for one
specific configuration in the degenerate family that was explicitly
known, apart from the trivial standard wall with $\chi = 0$.  In this
work  in Sect. 3 we construct the solutions for all configurations
belonging to this family. For arbitrary ratio of the coupling constants
$\rho \equiv \lambda/\alpha$ this solution is semi-explicit in the sense
that the trajectory in the field space is found in terms of elementary
functions, while the dependence of the fields on  $z$, although expressed in
quadratures, is not found analytically. An explicit dependence on $z$
can be found in terms of elementary functions for  two special values of
$\rho$: $\rho =1$ and $\rho =4$. The case of $\rho =1$ is, however,
trivial since after a $\pi/4$ rotation in the space of the fields
$(\phi, \, \chi)$ the model reduces to two fields not interacting with
each other. The remaining case of $\rho =4$ is quite non-trivial and
provides a whole family of new domain  wall solutions.

In four dimensions the requirement of renormalizability restricts the
form of the superpotential: it must be polynomial in fields of at most
third order. If we dimensionally reduce the theory to two dimensions,
the choice becomes infinitely richer. Any analytic function of $\Phi$
and $X$ can serve as a superpotential, without spoiling the
renormalizability of the two-dimensional model. We briefly discuss the
issue of the continuous degeneracy of the soliton solutions in this
setting (Sect. 2).

The domain wall configurations, when viewed as depending on the
Euclidean time $\tau$ rather than on the spatial coordinate $z$, are
known$^{\cite{brown}}$ to be the generating functions for amplitudes of
multiple production of bosons at  threshold by field operators. For the
model discussed here these are the amplitudes for production of  an
arbitrary number $n_\phi$ of the $\phi$ bosons and $k_\chi$ of the
$\chi$ bosons at the corresponding thresholds,
$$
A^{\phi}_{n \, k} \equiv
\langle n_\phi \, k_\chi | \phi(0) | 0 \rangle \,\,\,\mbox{and}\,\,\,
A^{\chi}_{n \, k}
\equiv \langle n_\phi \, k_\chi | \chi(0) | 0 \rangle\, .
$$
In Sect. 4 we use the relation between the domain wall profile and the
threshold production amplitudes in a twofold way: to point out a
constraint on  the solutions in the degenerate family stemming from the
fact that they generate the same set of amplitudes and to find the
multi-boson amplitudes at the tree level explicitly in the case of $\rho
=4$ where the explicit form of the solutions is available.

\section{Additional Integrals of Motion}

To begin with, we will discuss the occurrence of an additional integral
of motion in the simplest example (\ref{spot}). In this model the creek
equations have the form
$$
\dot \phi = \frac{\partial W}{\partial \phi }=\frac{m^2}{\lambda}
-\lambda\phi^2 -\alpha \chi^2\, ,
$$
\beq
\dot \chi = \frac{\partial W}{\partial \chi } = - 2\alpha \phi\chi\, .
\eeq
Let us introduce a ``dual" function
\beq
\tilde W(\Phi , X) =
X^{-\rho }\left(  \frac{m^2}{\lambda^2} - \Phi^2
-\frac{1}{\rho -2} X^2\right) \, ,
\label{dual}
\eeq
where
$$
\rho \equiv \frac{\lambda}{\alpha}  .
$$
The meaning of the word ``dual" will become clear shortly.
Equation (\ref{dual}) assumes that the parameter $\rho \neq 2$.
The case $\rho = 2$ is special and has to be treated separately.

The dual function has the property
\beq
\frac{\partial\tilde W}{\partial \phi_i}
= \varepsilon_{ij} S(\phi_i ) \frac{\partial W}{\partial \phi_j}
\label{defdu}
\eeq
where $\phi_{1,2}$ stands for $\{\phi , \chi\}$, $\varepsilon_{ij}$
is the antisymmetric tensor of the second rank, and $S $ is some
scalar function. In the model at hand
\beq
S(\Phi, X) = \frac{1}{\alpha}\frac{1}{X^{\rho +1}}\, .
\eeq

Now, if $z$ is interpreted as ``time", the dual function
$\tilde{W}$ is conserved along the trajectory. Indeed,
\beq
\dot{\tilde{W}} = \frac{\partial\tilde W}{\partial \phi_i}
\dot\phi_i = \frac{\partial\tilde W}{\partial \phi_i}
\frac{\partial W^\dagger}{\partial \phi_i^\dagger}
\, .
\label{van}
\eeq
For real solutions Eq. (\ref{defdu}) implies that the right-hand side of
Eq. (\ref{van}) vanishes.

Therefore, each particular real trajectory connecting the stationary
points 1 and 2 of the superpotential is characterized by the value of
the dual function on this trajectory. More exactly, dual functions
conserved along the trajectory can be introduced for all solutions
$\phi_k (z)$ with the constant, i.e. $z$ independent, phases of the
fields $\phi_k$. By an appropriate redefinition of the fields we can
obviously return to the real solutions.

In the general case of non-minimal models the superpotential (restricted
to real values of the superfields) is characterized by the gradient
lines and the level lines. The latter correspond to fixed values of $W$.
Two nets of lines -- gradient and level -- are locally orthogonal to
each other.  The level lines of the dual function are the gradient lines
of the superpotential while the gradient lines of the dual function are
the level lines of the superpotential. From this graphic interpretation
it is intuitively clear that a dual function $\tilde W$ must exist for
every $W$, although, unlike the minimal two-field model, it is not
always possible to find them analytically.  The points where $W$
(restricted to the real values of the scalar fields) develops maxima or
minima are the singular points of $\tilde W$.  The saddle points of $W$
are the saddle points of $\tilde W$.

\subsection{Solitons in 1+1 dimensions}

If in four dimensions the choice of the superpotential is severely
restricted by the requirement of renormalizabilty (only polynomials
which are at most cubic are allowed), in two dimensions any
superpotential leads to a sensible quantum theory. If one takes a
generalized Wess-Zumino model in four dimensions, with arbitrary number
of fields and an arbitrary superpotential, and dimensionally reduces it
to two dimensions, one arrives at $N=2$ supersymmetry in two dimensions.
In two dimensions the domain walls become solitons -- localized field
configurations with finite energy. After quantization they are to be
viewed as particles.  The $N=2$ supersymmetric theories with scalar
superfields in two dimensions were extensively studied in Ref.
\cite{vafa}.  The case which is of most interest to us, a continuous
family of degenerate solitons, seemingly was not discussed in this work.

Since we are not limited now to polynomial superpotentials we can
consider whole  families of models. Consider for definiteness
two-field models. Let us start from a certain model with a
superpotential $W$. One then can consider any other model with
a superpotential $W_{\rm NEW} = f(W)$ where $f$ is an arbitrary
function. If in the original model the dual function is known,
it remains the same for the whole family. Indeed, Eq. (\ref{defdu})
implies that
\beq
\frac{\partial\tilde W}{\partial \phi_i}
= \varepsilon_{ij} S_{\rm NEW}\frac{\partial W_{\rm NEW}}{\partial
\phi_j}\, , \,\,\,  S_{\rm NEW} = S\left( df/dW\right)^{-1}\, .
\label{defdunew}
\eeq
In other words, $\tilde W$ remains the integral of motion
for real trajectories in any model belonging to the given family.

The minimal two-field family, dimensionally reduced to $D=2$,
presents a simplest example where continuously degenerate soliton
solutions exist. Now one can easily provide with a plethora of
other interesting examples. For instance, one can consider
superpotentials which are bounded from above and from below for
real values of the superfields. In such models typically every soliton
will appear as a member of a degenerate family of solitons,
the degeneracy being unrelated to any external symmetry.
Generalizations of the sine-Gordon model fall into this category.
Consider, for example, the superpotential
\beq
W = - \sin\Phi -\sin X -\alpha (\sin\Phi )( \sin X )\, ,
\label{gsg}
\eeq
where $\alpha$ is a dimensional parameter, and all dimensional
parameters are set equal to unity. This superpotential is periodic in
$\Phi$ and $X$; for $\alpha = 0$ it describes two decoupled  fields
(each of them presents a supergeneralization of the sine-Gordon model).
If $\alpha\neq 0$ the fields $\Phi$ and $X$ start interacting with each
other. Inside the periodicity domain $0\le \Phi , X\le 2\pi$ the relief
of the superpotential $W$ is qualitatively similar to that of the
minimal two-field model: $-W$ has one maximum at $\Phi = X = \pi /2$,
one minimum at $\Phi = X = 3\pi /2$, and two saddle points at $\Phi =\pi
/2,\,\, X = 3\pi /2$ and $\Phi =3\pi /2,\,\, X = \pi /2$ (at least for
small values of $\alpha$). Any real trajectory (out of a continuous
family of trajectories) starting at the maximum leads to a minimum. The
only exceptions are two trajectories leading to the saddle points (the
exceptional. or basic solitons). The masses of two exceptional solitons
are $4+4\alpha$ and $4-4\alpha$, respectively. Continuously degenerate
solitons are bound states of two basic solitons, with mass 8, i.e. the
binding energy exactly vanishes.

Unlike the minimal two-field model all solutions in the model
(\ref{gsg}) have finite masses; there are no trajectories leading to
abysses.

The degeneracy is not lifted due to quantum corrections. This suggests
that in every model with continuously degenerate solitons there should
exist a dual description where the exceptional solitons (comprising the
degenerate ones) appear as decoupled (i.e. not interacting with each
other) particles from the very beginning.  This issue  deserves further
investigation.

\section{Solution for the Degenerate Walls}

Now we return to the minimal two-wall models (\ref{spot}). Our task is
to find the family of the wall trajectories (i.e. $\phi$ versus $\chi$
for every given wall solution and every allowed value of $\tilde W$).
Next we will find the wall solutions themselves, i.e. $\phi (z) $ and
$\chi (z)$.

\subsection{General case: arbitrary ratio $\rho \equiv
\lambda/\alpha$}

It is convenient to introduce dimensionless field variables $f$ and
$h$ as
$$
\phi={m \over \lambda} \, f\, , \,\,\, \chi= {m \over \sqrt{\lambda \,
\alpha}}
\, h\, ,
$$
and to set the mass parameter $m$ to one (it can be restored,
if needed, from dimension). The BPS-saturation equations (\ref{bpsr})
then take the form
\begin{eqnarray}
{df \over dz}&=&1-f^2-h^2 \nonumber \\
{dh \over dz}&=&-{2 \over\rho} \, f \, h~~.
\label{bpsm}
\end{eqnarray}
By eliminating the variable $z$ from these equations one finds the
equation for the trajectory in the $(f,h)$ plane,
\beq
{df \over dh}=-{\rho \over 2} \,{{1-f^2 - h^2} \over f \, h}~~.
\label{traj}
\eeq
The general solution of this equation can be written as
\beq
f^2=1-{ \rho \, h^2 \over {\rho-2}}- C \, h^\rho~,
\label{sg}
\eeq
where $C$ is an integration constant. It is connected with the integral
of motion, Eq. (\ref{dual}), by a simple proportionality relation,
$$
\tilde W = \left( \frac{m}{\lambda}\right)^{2-\rho}\, \rho^{-\rho /2}
C\, .
$$
The full trajectory runs from the point $(f,h)=(-1,0)$ in the ``distant
past" (i.e. $z=-\infty$) to the point $(f,h)=(1,0)$ in the ``distant
future" (i.e. $z=\infty$). The relation (\ref{sg}) determines it
piecewise: from $(f,h)=(-1,0)$ to $(f,h)=(0,h_{\rm max})$ and from
$(0,h_{\rm max})$ to $(1,0)$, where $h_{\rm max}$ is the maximal
amplitude of $h$ on the trajectory (see Fig. 1). For  the given value of
$C$ the value $h_{\rm max}$ is obviously determined from the solution of
Eq. (\ref{sg}) for $h$ with $f$ being set to zero,
\beq
\frac{\rho}{\rho -2} h_{\rm max}^2 + Ch_{\rm max}^\rho = 1\, .
\label{hmax}
\eeq
Using the symmetry under $h \leftrightarrow -h$ it is sufficient to
discuss only the trajectories with positive $h$.

The freedom  in $C$ is in fact limited by the condition that the fields
remain real along the whole trajectory. In this connection it should be
noted that the BPS-saturation equations (\ref{bps}) are intrinsically
non-analytic, thus the trajectory found for the real values of the
fields cannot be continued to the the complex values of the fields. The
requirement of the real trajectories translates into the requirement
that $h_{\rm max}$ is real positive and varies in the interval $[0 ,
1]$.  An elementary inspection of Eq. (\ref{hmax})  yields the allowed
domain for the constant $C$,
$$
C_* \le C \le +\infty\, , \,\,\,  C_* = \frac{2}{2-\rho}\, .
$$
One can readily see that the trajectory with $C=+\infty$ has $h(z)\equiv
0$ and, thus, reduces to the well-known wall solution in the  one-field
theory, $f(z)=\tanh z$. On the other hand the trajectory with $C$ equal
to the critical value $C=C_*$, has $h_{\rm max}=1$. Thus, it  in fact
describes two infinitely separated domain walls: one interpolating
between $(f,h)=(-1,0)$ and $(0,1)$ (the saddle point of $W$) and the
other between $(0,1)$ and $(1,0)$. The configurations with the
intermediate values of $C$ interpolate between these two extremes; in a
sense, they can be viewed as the solutions with the latter two walls at
a finite separation (see Fig. 2). Remarkably, the degeneracy of the
energy of the $C>C_*$ solutions  implies that the latter two walls do
not interact with each other.

\begin{figure}
\epsfxsize=9cm
\centerline{\epsfbox{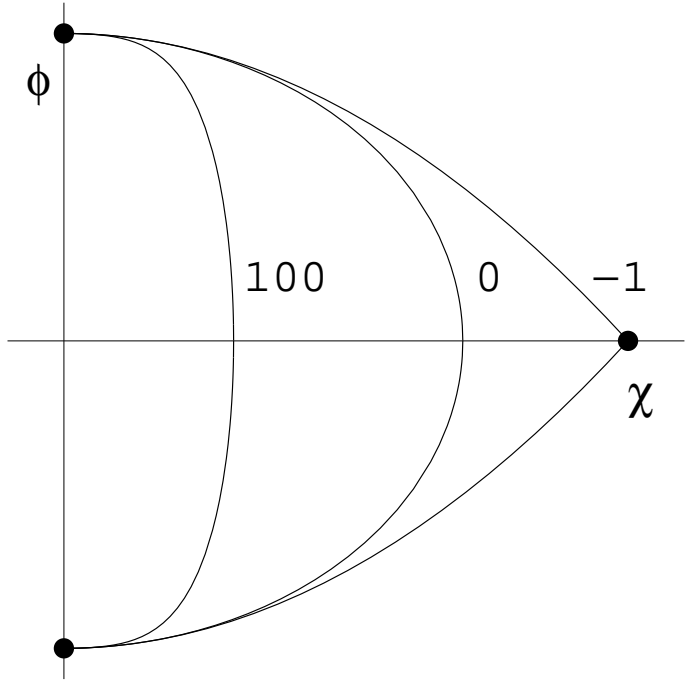}}
\caption{Few trajectories in the field space from the family of
degenerate
solutions (at $\rho=4$). The numbers represent the values of the
constant $C$
for the corresponding trajectory. The heavy dots show the positions
of
the
vacuum states.}
\end{figure}

\begin{figure}
\epsfxsize=14cm
\centerline{\epsfbox{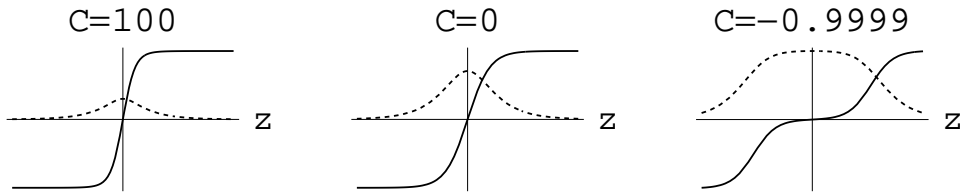}}
\caption{The profile of the fields $\phi$ (solid) and $\chi$ (dashed) as
a function of the spatial coordinate z for three values of $C$
($\rho=4$). It is seen that at $C$ approaching the critical value $C_* =
-1$, the profile separates into two walls.}
\end{figure}

If $\rho > 2$ the point $C=0$ belongs to the allowed interval. For $C=0$
(i.e. $\tilde W =0$) the trajectory takes an algebraic form. One can
immediately find an explicit domain wall solution, see Eq. (\ref{msol}).
As a matter of fact, it was obtained  previously$^{\cite{ms}}$.

It should be noted that the value $\rho =2$ presents a special case
because of the singularity in Eq. (\ref{sg}) at $\rho =2$. This
singularity can easily be resolved, however, either by considering the
limiting procedure in Eq. (\ref{sg}), or by solving the equation
(\ref{traj}) separately for $\rho =2$. The result is that the relation
(\ref{sg}) at $\rho =2$ reduces to

\beq
f^2=1+h^2 \, \left ( \ln h^2 - \tilde C \right )~~,
\label{ssg}
\eeq
where $\tilde C$ is another integration constant bound by the
condition $1\le \tilde C \le +\infty$.

The relation (\ref{sg}) gives the trajectory of the field configuration
in the $(f,h)$ plane (see Fig. 1). In order to find the coordinate
dependence of each of the fields, one can substitute $f$ as found from
Eq. (\ref{sg}) into the second of the equations (\ref{bpsm}) and then
obtain  the solution in an implicit form, ``in quadratures"
\beq
z= -{\rho \over 2} \int { dh \over h \, f(h) }~~.
\label{qua}
\eeq

\subsection{Explicit solution for $\rho=4$}

For arbitrary values of  $\rho$ there is no algebraic expression for
the integral in Eq. (\ref{qua}) in
terms of known functions. The exceptional cases are
$$
\rho ={1 \over 2},
\, {2\over 3},  \, 6,\,\,\,\mbox{ and}\,\,\,  8\, ,
$$
when the integral is expressed in terms of
elliptic functions, and
$$
\rho =1\,\,\,\mbox{and} \,\,\, \rho =4\, ,
$$
when the integral is elementary.

The elliptic cases are rather cumbersome, while the case of $\rho=1$ is
trivial: at $\rho=1$ the model considered here describes two fields
$\tilde \phi= (\phi+\chi)/\sqrt{2}$ and $\tilde \chi=
(\chi-\phi)/\sqrt{2}$ that do not interact with each other. For these
reasons we pursue here the explicit solution only for the exceptional
case of $\rho=4$.

If we choose this specific value,  $\rho=4$,  in Eqs. (\ref{sg}) and
(\ref{qua}), we readily find the explicit expressions for the fields $f$
and $h$ versus $z$,
\beq
f(z)={u^2-C-1 \over (u+1)^2+C}~~, ~~~~ h^2(z)={ 2 \, u \over
(u+1)^2+C}~~,
\label{sol}
\eeq
where $u=\exp (z-z_0)$ and $z_0$ is an arbitrary shift of the
coordinate $z$. (Clearly, the solution can be centered at $z=0$ so that
$f(0)=0$, if $z_0$ is chosen as $2 \, z_0 = - \ln (C+1)$.)
For $\rho=4$ the integration constant $C$ is bounded by the
condition
$$
-1 \le C \le +\infty\, .
$$
For completeness we also present the same solution in terms of the
fields $\phi(z)$ and $\chi(z)$, with  the normalization factors restored
(Fig. 2),
\beq
\phi(z)= {m \over \lambda} \, {e^{2\, m \, (z-z_0)}-C-1 \over { \left (
e^{ m \, (z-z_0)}+1 \right )^2 + C}}~~, ~~~~ \chi(z)= \pm {m \over
\lambda}
{2 \sqrt{2}\, e^{ m \, (z-z_0)/2} \over \sqrt{ \left ( e^{ m \,
(z-z_0)}+1
\right )^2 + C}}~.
\label{sol1}
\eeq
The constant $C$ plays the role of a collective coordinate. The
occurrence of another collective coordinate, $z_0$, is a trivial
consequence of the fact that the wall solution spontaneously breaks
the translational invariance of the original model in the $z$ direction.

At the same time, $C$ is unrelated to spontaneous breaking of any
symmetry of the model. It reflects the fact that the symmetry of the
particular solutions at hand  is higher than that of the model {\em per
se}.  As far as we know, Eq. (\ref{sol1}) presents the first explicit
example of a soliton family in the renormalizable $\phi^4$ theory with a
continuous degeneracy due to continuous deformations of the soliton
structure.

One can readily derive from the explicit solution above the limiting
cases discussed in Sect. 2.1. At the critical value $C=-1$ the
solution in Eq. (\ref{sol}) degenerates into
\beq
f={u \over u+2}~~, ~~~~ h^2={2 \over u+2}
\label{solc}
\eeq
and describes the domain wall connecting the vacua $(f,h)=(0,1)$ and
$(1,0)$. In order to get the wall connecting $(f,h)=(-1,0)$ with $(0,1)$
one can use the symmetry under $f \leftrightarrow -f$ and reverse the
sign of $f$ in the first equation in (\ref{solc}). In order to recover
the one-field solution in the limit $C \to +\infty$ one has to
accordingly adjust the coordinate shift $z_0$ as $\exp (-2 \, z_0) =
C+1$. Then in the limit $C \to +\infty$ one gets $h=0$ and $f= \tanh z$.

\section{Threshold multiparticle amplitudes}

In this section we will take advantage of the explicit wall family
solution found above in order to extract certain predictions for the
high-order behavior of the multiparticle amplitudes at thresholds. The
corresponding analysis for the one-field Wess-Zumino model was carried
out in Ref. \cite{cs}. In the one-field model there is little
distinction with the non-supersymetric case (for a review see Ref.
\cite{mv2}). The two-field model is much more interesting since it
reveals a new pattern.

In what follows we will need to know that in the vacuum with $\phi =
m/\lambda$ and $\chi = 0$ (from which our wall trajectory starts) the
mass of the $\phi$ quantum is equal to $2m$ while the mass of the $\chi$
quantum is equal to $2m/\rho $. In the case $\rho =4$ to be analyzed
below the mass of the $\chi$ quantum is equal to $m/2$. The same is
valid for the vacuum $\phi = m/\lambda\, , \,\,\, \chi = 0$, from which
the trajectories originate.

\subsection{An overview of the formalism}

The solutions to the field equations, in particular the domain wall
solutions, are directly related to multiparticle amplitudes, by virtue
of   the formalism developed by Brown$^{\cite{brown}}$ (for a more
recent review see Ref.\cite{mv2}). Being adapted to the present problem
of two fields the formalism is constructed as follows. Consider for
definiteness the amplitude  $\langle n, \,k |\phi(0)|0 \rangle$
describing the production by the field operator $\phi(x)$ of a
multiparticle state consisting of $n$ on-shell bosons of the field
$\phi$ with 4-momenta $p_a$ ($a=1, \ldots,n$) and $k$ on-shell bosons of
the field $\chi$ with 4-momenta $p_b$ ($b=1, \ldots, k$) in a vacuum $|
0 \rangle$ of the theory. According to the standard reduction formula
this amplitude is expressed through the response  of the system to
external sources $\rho_\phi(x)$ and $\rho_\chi(x)$, coupled to the
corresponding fields as $\rho_\phi \, \phi+\rho_\chi \, \chi$ in the
Lagrangian,
\begin{eqnarray}
&&\langle n, \,k |\phi(0)|0 \rangle=\left [ \prod_{a=1}^n \lim_{p_a^2
\to
m_\phi^2}
\int d^4 x_a~ e^{i p_a x_a} (m_\phi^2-
 p_a^2)~ {\delta \over {\delta \rho_\phi(x_a)}}
\right ] \times \nonumber \\
&& \left [ \prod_{b=1}^k \lim_{p_b^2 \to m_\chi^2}
\int d^4 x_b~ e^{i p_b x_b} (m_\chi^2-
 p_b^2)~ {\delta \over {\delta \rho_\chi(x_b)}}
\right ]
\left . \langle 0_{out}|\phi(x)|0_{in}\rangle ^{\rho_\phi, \, \rho_\chi}
\right|_{\rho_\phi=0, \, \rho_\chi=0}~~,
\label{lsz}
\end{eqnarray}
where $m_\phi$ and $m_\chi$ are the masses of the respective
bosons in the vacuum considered.

The classical response, i.e. the classical solution of the field
equations in the presence of the sources generates, through Eq.
(\ref{lsz}), the tree-level amplitudes, which we will be mainly
concerned with here. Moreover, as will be seen, the configurations, of
the type of  the domain walls, depending on only one variable, are
related to multiparticle production exactly at the threshold, i.e. at
the spatial momenta of the produced particles exactly equal to zero. In
this situation it is sufficient to consider the response of the system
in Eq. (\ref{lsz}) to spatially uniform time-dependent sources,
$$
\rho_\phi(t)= \rho_\phi(\omega_\phi) \,
e^{i
\omega_\phi t}\, , \,\,\, \rho_\chi(t)= \rho_\chi(\omega_\chi) \, e^{i
\omega_\chi t}\, ,
$$
and take in the very end the on-shell limit in Eq. (\ref{lsz}) by
tending $\omega_\phi$ to $m_\phi$ and $\omega_\chi \to m_\chi$. The
spatial integrals in Eq. (\ref{lsz}) then give the normalization spatial
volume, conventionally set to one, while the time dependence with the
fixed functional form of the sources implies that the propagator factors
and the functional derivatives enter in the combination
\begin{eqnarray}
&&(m_\phi^2-p_a^2){\delta \over {\delta \rho_\phi(x_a)}} \to
(m_\phi^2-\omega_\phi^2) {\delta
\over {\delta \rho_\phi(t)}}={\delta \over {\delta a_\phi(t)}}~~,
\nonumber \\
&&(m_\chi^2-p_b^2){\delta \over {\delta \rho_\chi(x_b)}} \to
(m_\chi^2-\omega_\chi^2) {\delta
\over {\delta \rho_\chi(t)}}={\delta \over {\delta a_\chi(t)}}~~,
\label{at}
\end{eqnarray}
where
\beq
a_\phi(t)={{\rho_\phi(\omega_\phi)\, e^{i \omega_\phi t}} \over
{m_\phi^2-i \epsilon -\omega_\phi^2}}~~, ~~~~
a_\chi(t)={{\rho_\chi(\omega_\chi)\, e^{i \omega_\chi t}} \over
{m_\chi^2-i \epsilon -\omega_\chi^2}}~~
\label{adef}
\eeq
coincide with the response of free fields to the external sources. For
finite amplitudes of the sources the response is singular in the
on-shell limit $\omega_\phi \to m_\phi$, $\omega_\chi \to m_\chi$.
Therefore, following Brown$^{\cite{brown}}$, the amplitudes of the
sources should be taken to zero in this limit, so that the factors
$a_\phi(t)$ and $a_\chi(t)$ are finite,
$$
a_\phi(t) \to a_\phi \, e^{i m_\phi t}\, , \,\,\, a_\chi(t) \to a_\chi
\,
e^{i m_\chi t}\, .
$$

Thus, for the purpose of calculating the multiparticle amplitudes at the
tree level one looks for a solution of the classical field equations
with no sources. The only information about the sources left from the
above limiting procedure is that the sources drive only positive
frequencies in the fields, thus the condition for the sought solution is
that it should contain only the positive frequency part with all
harmonics being multiples of $e^{i m_\phi t}$ and $e^{i m_\chi t}$. The
latter condition is equivalent to requiring that the solution for the
fields goes to the classical vacuum at  infinity in the Euclidean time
$\tau={\rm Im} \, t \to +\infty$. The multiparticle amplitudes are then
given by the derivatives of the solution $\phi(t)$,
\beq
A_{n, \, k}^\phi \equiv \langle n, \, k | \phi(0) |0 \rangle = \left (
{\partial \over \partial a_\phi(t)} \right ) ^n \, \left ( {\partial
\over
\partial a_\chi(t)} \right ) ^k \, \phi(t) \left. \right |_{a_\phi=0, \,
a_\chi=0}~~.
\label{ap}
\eeq
Since the equations for the fields $\phi$ and $\chi$ are coupled, one
simultaneously finds the solution for the field $\chi$ and, thus, the
amplitudes for the multi-boson production by $\chi$,
\beq
A_{n, \, k}^\chi \equiv \langle n, \, k | \chi(0) |0 \rangle = \left (
{\partial \over \partial a_\phi(t)} \right ) ^n \, \left ( {\partial
\over
\partial a_\chi(t)} \right ) ^k \, \chi(t) \left. \right |_{a_\phi=0, \,
a_\chi=0}~~.
\label{ac}
\eeq

The operational procedure for calculating the amplitudes is therefore as
follows. First, one obtains the solution of the Euclidean classical
field equations depending only on  time $\tau$ and approaching, at $\tau
\to +\infty$, the vacuum state $(\phi_0,\chi_0)$ in which the scattering
theory is considered. Then, the  solution is expanded in the harmonics
of $e^{-m_\phi \, \tau}$ and $e^{-m_\chi \, \tau}$,
\beq
\phi(\tau)=\sum_{n=0, \, k=0}^\infty \, F_{n,\,k}\, e^{-(n \, m_\phi+ k
\,
m_\chi)\, \tau}~~, ~~~~ \chi(\tau)=\sum_{n=0, \, k=0}^\infty \,
H_{n,\,k}\, e^{-(n \, m_\phi+ k \, m_\chi)\, \tau}~~,
\label{fh}
\eeq
where $F_{n,\,k}$ and $H_{n,\,k}$ are the coefficients of the
expansion.

Note that
$$
F_{0, \, 0}=\phi_0\,\,\,\mbox{and}\,\,\, H_{0, \, 0}=\chi_0\, ,
$$
while the
coefficients of the appropriate first harmonics are identified as the
described above factors $a_\phi$ and $a_\chi$,
$$
F_{1, \,0}=a_\phi\, , \,\,\,
H_{0,\, 1}=a_\chi\, .
$$
 Then, according to Eqs. (\ref{ap}) and
(\ref{ac}), the amplitudes are expressed as
\beq
A_{n, \, k}^\phi = n! \, k! \, {F_{n, \, k} \over F_{1, \, 0}^n \, H_{0,
\, 1}^k }~~, ~~~~
A_{n, \, k}^\chi = n! \, k! \, {H_{n, \, k} \over F_{1, \, 0}^n \, H_{0,
\, 1}^k }~~.
\label{afh}
\eeq

Before closing this discussion of the general formalism we would like to
emphasize that the latter equations can be also viewed as a constraint
on any solution approaching a vacuum state $(\phi_0, \, \chi_0)$ at
$\tau \to +\infty$. Namely, if up to the appropriate linear terms the
fields behave as
$$
\phi(\tau) = \phi_0 + a \, e^{-m_\phi \,t}+ \ldots\, , \,\,\, \chi(\tau)
=
\chi_0 + b \, e^{-m_\chi \,t}+ \ldots\, ,
$$
then the subsequent terms in the expansion of the fields are fully
determined in terms of $a$,  $b$ and the  fixed set of the amplitudes
$A^\phi$ and $A^\chi$,
\beq
F_{n, \, k}={ A^\phi_{n, \, k} \, a^n \, b^k \over n! \, k!}~~, ~~~~
H_{n, \, k}={ A^\chi_{n, \, k} \, a^n \, b^k \over n! \, k!}~~.
\label{fha}
\eeq

Furthermore, the absolute normalization of the coefficients $a$ and $b$
is rather a matter of convention. Indeed, under a shift of $\tau$,
$$
\tau \to \tau-\tau_0
$$
these coefficients change as
$$
a \to a \, e^{m_\phi \,
\tau_0}\, , \,\,\,  b \to b \, e^{m_\chi \, \tau_0}\, .
$$
Thus, the only parameter that distinguishes between essentially
different solutions approaching the same vacuum state is the ratio
$$
c=\frac{a}{b^{m_\phi/m_\chi}}\, .
$$
Therefore,  in a general two-field theory a family of solutions
approaching a vacuum state at $\tau \to +\infty$ is parametrized by a
single  parameter $c$. This parameter is in one-to-one correspondence
with the integration constant $C$, or the value of the dual function
$\tilde W$ on the trajectory.  The coefficients  of the expansion
(\ref{fh}) are then fixed by the multiparticle amplitudes.

\subsection{Multiparticle amplitudes in the supersymmetric model}

The domain wall solutions discussed in Sect. 2 can be directly applied
to calculating the multiparticle amplitudes. To this end one should
consider the fields depending on the Euclidean time $\tau$ rather than
on the spatial coordinate $z$. Since this amounts to a trivial
relabelling of the variable, we retain here the notation $z$ for the
variable. We also use the notation $a=F_{1,\,0}$ and $b=H_{0, \, 1}$.
Every BPS-saturated solution from the family under consideration
approaches at $z \to +\infty$ the vacuum at $(\phi,\,
\chi)=(m/\lambda,\, 0)$. Thus, this is the vacuum state in which the
multiparticle amplitudes are generated by the solutions. Remember that
the masses of the particles in this vacuum are expressed in terms of the
parameters of the model as $m_\phi=2 \, m$, $m_\chi = 2 \, (\alpha /
\lambda )  m =  m_\phi /\rho $.

In the case of  arbitrary ratio $\rho$  one can use Eq. (\ref{sg}) to
obtain a relation between the coefficients $a$, $b$ and the constant
$C$. Indeed, at $z \to +\infty$ the field $\chi(z)$ goes to zero as
$$
\chi(z)=b \, e^{-m_\chi \, z} +\ldots\, ,
$$
corresponding in the dimensionless variables to
$$
h(z) = { \sqrt{\lambda
\alpha} \, b \over m} \, e^{-2 z/\rho }\, .
$$
 The linear in $e^{-2 \, z}$
harmonics in $f(z)$ arises from the term in Eq. (\ref{sg}) with the
constant $C$,
from where one finds the coefficient $a$ of the linear in $e^{-m_\phi
\, z}$ harmonics in $\phi(z)$ ($\phi(z)={m \over \lambda} + a \, e^{-
m_\phi
\, z}+ \ldots$) as
\beq
a=-{C \over 2} \, {m \over \lambda} \, \left ( { \sqrt{\lambda \alpha}
\, b \over m} \right )^\rho~~.
\label{abc}
\eeq

In connection with the derivation of the latter relation it should be
noted that for $\rho  > 2$ the harmonics with $\exp (-m_\phi \, z)$ is
not the leading one in the field $\phi(z)$ at large $z$ because of the
presence of the second harmonics with the mass of $\chi$: $F_{0,\,2} \,
\exp (-2 \, m_\chi \, z)$. Thus, the presented derivation, strictly
speaking, is justified only at $\rho < 2$. However, the relation
(\ref{abc}) is also applicable at $\rho > 2$ since the fields and the
coefficients of their expansion are analytic functions of the couplings,
and the relation (\ref{abc}) can be analytically continued from $\rho <
2$ to the domain $\rho > 2$.

It can be also noted that the solution in Eq. (\ref{msol}) with $C=0$
has $a=0$, according to Eq. (\ref{abc}). Thus, it generates only the
amplitudes of multiple production of the bosons of the field $\chi$ by
either the operator $\chi(0)$ or $\phi(0)$. For this reason it expands
in the harmonics determined only by the mass of $\chi$.

Furthermore, for a rational ratio $\rho$ the masses $m_\phi$ and
$m_\chi$ are also in a rational proportion. Thus if only the $z$
dependence of the fields were known, there would be an ambiguity, at
least in some harmonics, in separation between  the production of the
$\chi$ bosons and the $\phi$ bosons. However, this ambiguity is resolved
if the dependence of the solution on $C$ is known, by using Eq.
(\ref{abc}), which shows that the constant $C$ serves as a ``tag" for a
$\phi$ boson. The power of $C$ in the given  harmonics gives the number
of the $\phi$ bosons in the amplitude generated by this harmonics.

We illustrate this method for our explicit solution in the case of
$\rho=4$ and we also find explicitly the amplitudes $A^\phi_{n, \, k}$
and $A^\chi_{n, \, k}$ in this case.

Setting for definiteness $z_0=0$ in the explicit solution in
Eq. (\ref{sol1}), we find the coefficient $b$ determining the rate of
approach of the field $\chi(z)$ to its vacuum value (zero),
\beq
b={ 2 \, \sqrt{2} \, m \over \lambda}~~.
\label{b4}
\eeq
Furthermore, using the relation (\ref{abc}) at $\rho=4$ we also
obtain the coefficient $a$,
\beq
a=-2 \, C \, {m \over \lambda}~~.
\label{a4}
\eeq
We then expand the expressions for the fields in Eq. (\ref{sol1}) in
powers of $C$ and, finally, each term of this expansion in powers of $e^{-m \,
z}$. In this way we get
\begin{eqnarray}
&&\phi(z)=\sum_{n=0,\, k=0}^\infty {\cal F}_{n, \, k} \, C^n \, e^{-
\left
( 2 \, n + {k \over 2} \right ) \,m \, z}~~, \nonumber \\
&&\chi(z)=\sum_{n=0,\, k=0}^\infty {\cal H}_{n, \, k} \, C^n \, e^{-
\left
(2 \, n + {k \over 2} \right ) \, m \, z}~~,
\label{calfh}
\end{eqnarray}
where the coefficients ${\cal F}_{n, \, k}$ (${\cal H}_{n, \, k}$) are
non-zero for even (odd) $k$, as is expected from the symmetry of
the  model at hand. The explicit expressions for these coefficients are
\begin{eqnarray}
&&{\cal F}_{n, \, k}={2 \, m \over \lambda} \, (-1)^{n + k/2} \,{
(2n+k/2)! \over (2n)! \, (k/2)!}~~,~~~~ (n+k >0)~~, \nonumber \\
&&{\cal H}_{n, \, k}= {2 \, \sqrt{2} m \over \lambda} \,
(-1)^{n+(k-1)/2}
\, { \Gamma(n+1/2) \, \left [ 2n + (k-1)/2 \right ] ! \over
\sqrt{\pi} \, n! \, (2n)! \, \left [ (k-1)/2 \right ] !}~~ .
\label{fh1}
\end{eqnarray}
The combinations ${\cal F}_{n, \, k} \, C^n$ and ${\cal H}_{n, \, k} \,
C^n$ are identified as respectively the coefficients $F_{n, \, k}$ and
$H_{n, \, k}$ in the general expansion of Eq. (\ref{fh}). The latter
coefficients are related to the multiparticle amplitudes as given by
Eq. (\ref{fha}). Using the explicit expressions in  Eqs. (\ref{b4})
and (\ref{a4}) for $b$ and $a$, one arrives at  the relation
between the amplitudes and the found coefficients ${\cal F}_{n, \, k}$
and ${\cal H}_{n, \, k}$. Namely,
\begin{eqnarray}
&&A^\phi_{n, \, k}= (-1)^n \, n! \, k! \, \left ( {\lambda \over m}
\right
)^{n+k} \, {{\cal F}_{n, \, k} \over 2^{n+3k/2}}=
(-1)^{k/2} \, \left ( {\lambda \over m} \right )^{n+k-1} \, { n! \, k!
\,
(2n+k/2)! \over 2^{n+3k/2-1} \, (2n)! \, (k/2)!}~~, \nonumber \\
&&A^\chi_{n, \, k}= (-1)^n \, n! \, k! \, \left ( {\lambda \over m}
\right
)^{n+k} \, {{\cal H}_{n, \, k} \over 2^{n+3k/2}} \nonumber \\
&&=
(-1)^{(k-1)/2} \, \left ( {\lambda \over m} \right )^{n+k-1} \,
{ k! \, \Gamma(n+1/2) \, \left [ 2n + (k-1)/2 \right ] ! \over
\sqrt{\pi}
\, 2^{n+3\, (k-1)/2} \, (2n)! \, \left [ (k-1)/2 \right ] !}~~ .
\label{resa}
\end{eqnarray}

This concludes our calculation of the  threshold amplitudes in the
minimal two-filed model.

\subsection{New zeros}
\begin{figure}
\epsfxsize=8cm
\centerline{\epsfbox{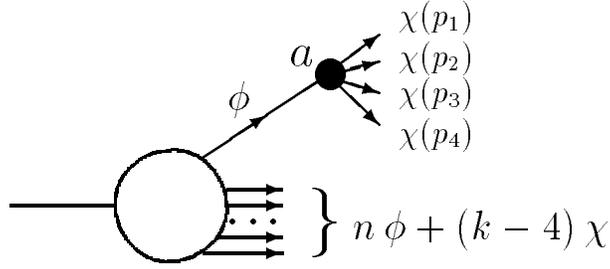}}
\caption{A graph with four $\chi$ bosons originating from a single
virtual $\phi$, which is singular at  threshold if $m_\phi/m_\chi=4$
unless the scattering amplitude $a$
(filled circle) vanishes at  threshold. Both the open and the filled
circles represent the sum of tree graphs.}
\end{figure}

A remarkable property of our  result  is that the amplitudes are
finite even though a state of four $\chi$ bosons at  threshold is
degenerate in energy with one $\phi$ boson. In other words, any graph,
where four final $\chi$ bosons with the four-momenta $p_1, \, p_2, \,
p_3$ and $p_4$ originate from a single line of $\phi$ (see Fig. 3),
contains the factor
\beq
{a_{1, \, 4} (p_1, \, p_2, \, p_3, \, p_4) \over (p_1+p_2+p_3+p_4)^2 -
m_\phi^2}~~.
\label{one4}
\eeq
Here $a_{1, \, 4} (p_1, \, p_2, \, p_3, \, p_4)$ is the conventional
Feynman scattering amplitude for the process $\phi \to 4 \, \chi$. At
threshold the denominator in Eq. (\ref{one4}) goes to zero, and the
graph becomes singular unless the Feynman amplitude $a_{1, \, 4} (p_1,
\, p_2, \, p_3, \, p_4)$ also vanishes when all four momenta are at
threshold. The latter cancellation indeed takes place in the model
considered here. This can be seen by examining the amplitude $A^\phi_{1,
\, 4}$, which exactly is the threshold limit of the expression in Eq.
(\ref{one4}). Indeed, the  amplitudes $A^\phi$ and $A^\chi$ considered
here are the matrix elements of the field operators in  Eqs. (\ref{ap})
and (\ref{ac}). These matrix elements have the propagators of the final
on-shell bosons amputated, but the propagator of the incoming virtual
field is not amputated and remains included in the corresponding
amplitude $A$. The conventional Feynman scattering amplitude is thus
obtained by multiplying the amplitude $A$ by the inverse propagator of
the incoming line. In the case of the  process $\phi \to 4 \, \chi$ the
inverse propagator of the incoming $\phi$ is vanishing at threshold of
four $\chi$. Thus, the Feynman scattering amplitude also vanishes.
Clearly, this cancellation can be also verified by an explicit
calculation of the tree Feynman graphs.

This cancellation can be extended to a general case of an arbitrary even
integer value of $\rho$, with the exception of $\rho=2$. In the latter
case the exponential behavior of $\chi$ at $z \to +\infty$ generates a
non-exponential dependence of $\phi$ through the logarithm in Eq.
(\ref{ssg}), which implies that in this case a resonance between the
degenerate states does take place. For all other values of $\rho$ the
coefficients in the expansion of the type as in Eq. (\ref{fh}) can be
constructed by iterations and are non-singular. This means that for the
values of $\rho$ where the resonance could potentially occur, i.e. $\phi
\to \rho \, \chi$ for even integer $\rho$, it actually does not take
place due to vanishing of the corresponding Feynman scattering
amplitude.

\section{Conclusions}

In summary,  in a class of supersymmetric models with the continuously
degenerate family of BPS domain walls (with real trajectories) an
additional integral of motion is observed. The occurrence of this
integral allowed us to find a generic solution from the family in
quadratures, while for a specific ratio of the coupling constants the
whole wall family is obtained in the closed form in terms of elementary
functions. We then further utilize the result for deriving the
multiparticle amplitudes at threshold in the minimal two-field
Wess-Zumino model. The threshold amplitudes are calculated in a closed
form for $\rho=4$. In the course of the calculation we have found an
unexpected cancellation of the tree graphs for the Feynman amplitude of
the process $\phi \to 4 \, \chi$ at the threshold, due to which
cancellation the multiparticle threshold amplitudes are finite. We also
conclude that the same cancellation takes place for the process $\phi
\to \rho \, \chi$ at arbitrary even integer $\rho$, except $\rho=2$. The
relation of this cancellation to additional integrals of motion is yet
to be studied. It can be also noted that the nullification of the
amplitudes is somewhat reminiscent of the general property of
nullification$^{\cite{mv3,mv4}}$ for the on-shell processes $2 \to many$
at the threshold in scalar theories.

Interesting phenomena occur when the models at hand are
dimensionally reduced to D=2. The two-dimensional theories thus
obtained have extended supersymmetry, $N=2$. A continuous
degeneracy of the soliton family (persisting with all quantum
corrections included) reflecting the possibility of the continuous
deformations of the solution profile can be seemingly interpreted in
this case as the existence of decoupled ``basic solitons".  Revealing
these decoupled basic solitons in an explicit form and studying their
properties is an obvious next problem to be dealt with in the given
range of questions.

\vspace{0.5cm}

{\bf Acknowledgments}: \hspace{0.3cm}

This work was supported in part by DOE under the grant number
DE-FG02-94ER40823.


\end{document}